\documentclass{nature}
\usepackage{graphicx}
\usepackage[normalem]{ulem}
\usepackage{amsmath}
\usepackage{color}
\usepackage{bm}
\usepackage{braket}
\usepackage{xcolor,cancel}
\usepackage{setspace}
\usepackage{hyperref}
\hypersetup{urlcolor=blue, colorlinks=false} 

\renewcommand\Re{\operatorname{Re}}
\definecolor{orange}{rgb}{1,0.5,0}

\title{Magnetic oscillations induced by phonons in non-magnetic materials}

\author{Idoia G. Gurtubay$^{1,2}$, Aitzol Iturbe-Beristain$^{1,2}$, Asier Eiguren$^{1,2}$}

\begin{document}

\maketitle

\begin{affiliations}
 \item Condensed Matter Physics Department, Science and Technology 
Faculty, University of the Basque Country UPV/EHU, PK 644, E-48080 Bilbao, Basque Country, Spain.
 \item Donostia International Physics Center (DIPC), Paseo Manuel de Lardizabal 4, E-20018 Donostia-San Sebasti\'{a}n, 
Spain.
Correspondence and requests for materials should be addressed to A.E. (email:\href{mailto:asier.eiguren@ehu.eus}{asier.eiguren@ehu.eus})
\end{affiliations}

\begin{abstract}
An unexpected finding two decades ago 
demonstrated that Shockley electron states in noble metal surfaces are spin-polarized,
forming a circulating spin texture in reciprocal space.
The fundamental role played by the spin degree of freedom was then revealed, even for a non-magnetic system, whenever
the spin-orbit coupling was present with some strength. 
Here we demonstrate that similarly to electrons in the presence of spin-orbit coupling, the 
propagating vibrational modes are also accompanied by a well-defined 
magnetic oscillation even in non-magnetic materials.
Although this effect is illustrated by considering a single layer of the WSe$_2$ dichalogenide, the
phenomenon is completely general and valid for any non-magnetic material with spin-orbit coupling.
The emerging phonon-induced magnetic oscillation  acts as an additional effective
flipping mechanism for the electron spin and its implications in the transport and scattering properties of the 
material are evident and profound.
\end{abstract}
%
\section*{Introduction}
In materials science, the most conventional point of view is to assume that propagating
 vibrational collective modes (phonons) 
are not associated with any magnetic property if the material itself is non-magnetic. 
In a magnetic material, however, it is natural to expect a magnetic oscillation associated to a phonon mode. 
%
Phonons are commonly understood as sinusoidal patterns of atomic displacements which couple to  
electron states by the scalar potential induced  by these atomic displacements. 
Electrons have a well defined spin-polarization
under spin-orbit interaction,
but when time reversal symmetry applies the spin polarizations for 
opposite momenta cancel each other and the material results to be non-magnetic. 
There is a clear evidence that the
lattice thermal conductivity of diamagnetic materials couples to external magnetic fields \cite{phdiamag}, 
which in principle might seem contrary to the idea that phonons do not have any associated magnetic property.
Here we show that
similarly to electrons, phonons are also accompanied by an  
induced effective magnetic oscillation when spin-orbit coupling is present even for non-magnetic materials. 
%

\section*{Results}

In a solid with spin-orbit coupling, electron states are described by two-component spinor wave functions for each  
$\mathbf k$ point spanning the Brillouin zone (BZ),
$\Psi_{{\bf k},i}({\bf r})= \left(\begin{array}{lcr}\!u^{\uparrow}_{{\bf k},i}({\bf r})\\
u^{\downarrow}_{{\bf k},i}({\bf r}) \end{array}\!\!\right) e^{ i {\bf k} {\bf r}}$. 
Each electron state has an associated spin-polarization 
defined as the expectation value of the Pauli vector 
$\mathbf{m}_{{\bf k},i}({\bf r}) = 
\bra{\Psi_{{\bf k},i}} {\boldsymbol \sigma} \ket{\Psi_{{\bf k},i}}$. 
Time reversed Kramers pairs at $\mathbf k$ and $-\mathbf k$ have opposite spin-polarizations
 that exactly cancel out when integrated, which therefore implies 
no magnetism. 
However, the electron spin-polarization is a crucial physical 
magnitude in many non-magnetic systems with spin-orbit coupling, 
one of the most outstanding being probably its role in the topological properties of matter.

A phonon excitation consists in a sinusoidal displacement of atoms 
and an induced (almost) static response of the electron gas 
which tries to weaken or screen out the electric perturbation generated by these displacements.
 Therefore, the question of whether
a phonon perturbation creates a magnetic oscillation could be suitably treated considering a generalized 
dielectric theory of dimension $4 \times 4$ mixing the magnetic and electric components of 
the potential \cite{JonL2017}. 
An alternative and more transparent way to see whether an overall magnetic property emerges is 
 to consider the effect of the perturbation on each electron spinor
wave function and then integrate over the BZ.
The key point is that if a phonon is excited with momentum ${\bf q}$ 
 it couples  differently  with the electrons at ${\bf k}$ and $-{\bf k}$,
 the result being that the spin-polarization
of electrons with time reversed momenta do not cancel each other. 
Under these conditions the balance of the electron spin-polarizations within the BZ is broken 
and the BZ integral gives a finite value  and, therefore, a net real-space 
magnetic oscillation with the same wave number as the phonon ${\bf q}$.

Let us focus on a single frozen phonon-mode ($\nu$) with momentum ${\bf q}$ which  
produces a perturbation on both components of the periodic part of the electron spinor wave functions, 
$\delta^{{\bf q},\nu} u^{\sigma}_{{\bf k},i}({\bf r})$, where $\delta^{{\bf q},\nu}$ denotes the self-consistent
variation in the context of density-functional perturbation theory \cite{Baroni2001,DalCorso2007}
and $u^{\sigma}_{{\bf k},i}({\bf r})$ is the periodic part of each spinor component. 
The unit cell periodic part of the amplitude of the frozen charge-spin density wave is obtained
by integrating the contributions from all occupied electron states (see Supplementary Note 1)
\begin{eqnarray} \label{eq:n2x2_tilde}
\delta \tilde{n}_{{\bf q},\nu}^{\sigma,\sigma'}({\bf r}) &=& 
\sum^{{\mathrm{occ}}}_{{\bf k},i}  \langle u^{\sigma'}_{{\bf k},i} ({\bf r}) \ket{ \delta^{{\bf q},\nu} u^{\sigma}_{{\bf k},i}({\bf r})  }.
\end{eqnarray}
Therefore,
$\delta \tilde{n}_{{\bf q},\nu}^{\sigma,\sigma'}({\bf r}) e^{i \bf {q}  \bf {r}}$  would represent 
the complete oscillation wave in real space.
The absence 
of a magnetic component accompanying a phonon mode would require  the off-diagonal elements
 ($\sigma\neq\sigma'$) to be zero and that 
both diagonal components are equal to each other,
 which, in general,  are conditions only fulfilled at the $\Gamma$ 
point (${\bf q=0}$).
 Writing the $2 \times 2$ charge-spin matrix of equation~(\ref{eq:n2x2_tilde}) 
in terms of Pauli matrices 
allows to explicitly  distinguish the electronic charge,
$\delta\tilde{n}_{{\bf q},\nu}^0({\bf r})$,
and the magnetic components, 
$\delta{\bf \tilde{M}}_{{\bf q},\nu}({\bf r})=
 (\delta\tilde{n}_{{\bf q},\nu}^1({\bf r}), 
  \delta\tilde{n}_{{\bf q},\nu}^2({\bf r}),
  \delta\tilde{n}_{{\bf q},\nu}^3({\bf r}))$: 
\begin{eqnarray}\label{eq:n1x4_tilde}
\nonumber
\left\{ \delta \tilde{n}_{{\bf q},\nu}^{\sigma,\sigma'}({\bf r}) \right\} \rightarrow 
\delta{\bf \tilde{n}}_{{\bf q},\nu}({\bf r}) &=& 
\delta\tilde{n}_{{\bf q},\nu}^0({\bf r}) \sigma_0 + \delta\tilde{n}_{{\bf q},\nu}^1({\bf r}) \sigma_1 + 
\delta\tilde{n}_{{\bf q},\nu}^2({\bf r}) \sigma_2 + \delta\tilde{n}_{{\bf q},\nu}^3({\bf r}) \sigma_3
\\
&=&  \delta\tilde{n}^0_{{\bf q},\nu}({\bf r}) \sigma_0+
\delta{\bf \tilde{M}}_{{\bf q},\nu}({\bf r}) \boldsymbol{\sigma}.
\end{eqnarray}

In real space, the time-dependent charge-spin field is 
given by the real part of the above frozen complex amplitudes 
when accounting for the classical motion of atoms. For a single phonon mode $({\bf q},\nu)$ 
of energy $\omega_{{\bf q},\nu}$ we have
\begin{eqnarray}
\nonumber
\delta{\bf n}_{{\bf q},\nu}({\bf r},t) &=&  \  {\Re} \left [  \left (  
  \delta\tilde{n}_{{\bf q},\nu}^0({\bf r}) \sigma_0+
	\delta{\bf \tilde{M}}_{{\bf q},\nu}({\bf r}) \boldsymbol{\sigma}\right ) {e^{i \left ( {\bf q} {\bf r}  - \omega_{{\bf q},\nu} t \right)} }\right ].
\\
\nonumber
&=&  \delta n_{{\bf q},\nu}^0({\bf r},t) \sigma_0+
\delta{\bf M}_{{\bf q},\nu}({\bf r},t) \boldsymbol{\sigma},
\end{eqnarray}
concluding that the appearance of an induced spin-density (or magnetization indistinctly)  
$(\delta{\bf \tilde{M}}_{{\bf q},\nu}({\bf r}))$
is completely general for crystals with spin-orbit coupling since the
 only requirement is a non-trivial pattern 
of the spin-polarization within the BZ associated to the absence of inversion symmetry~\cite{LaShell96}. 
The phonon modes break the symmetry of the BZ
in a way that the electron spin-polarization is modulated within the BZ  producing 
a net spin accumulation. 
The similarity of the phonon magnetism and the electron spin-polarization with 
time-reversal symmetry (no net magnetism) 
is strengthened by the fact that time-reversed phonon momenta  give strictly opposite 
magnetization exactly in the same way as for electrons  
$\delta{\bf M}_{{\bf q},\nu}({\bf r}, t)=-\delta{\bf M}_{{\bf -q},\nu}({\bf r}, t)$.

The above description of the spin-charge field induced by phonons is completely classical and focussed on
a single phonon with a fixed momentum. Therefore, physically $\delta{\bf M}_{{\bf q},\nu}({\bf r}, t)$ would be
the time dependent magnetization associated with a single coherent phonon mode. 
In general the vector field defined by $\delta{\bf M}_{{\bf q},\nu}({\bf r}, t)$ shows 
an interesting real-space and time dependent non-collinear pattern, which depends also on 
the particular atomic displacements (polarization vectors) associated with each phonon branch. 
Actually, it is the motion of the W atom, 
i.e.  the atom for which the spin-orbit interaction is dominant,
 which determines the direction of the magnetization.
For instance, for ${\bf q}$=${\bf K}$ and for a mode in which Se atoms rotate clockwise with opposite phase
in the plane of the surface ($x$-$y$ plane)  and W atoms vibrate in the
 perpendicular direction ($z$) to the surface, a net circularly 
polarized induced magnetization appears in the  surface plane  around the W atoms
 (Figs.~{\ref{fig:K_mod02}}{\bf a}-{\bf c} and Supplementary Movie 1). 
 However, for the same phonon propagation vector  ${\bf q}$=${\bf K}$, if   
W atoms rotate clockwise around their equilibrium positions
 in the plane of the surface, 
the net magnetization shows along the perpendicular direction to the plane
(Figs.~{\ref{fig:K_mod03}}{\bf a}-{\bf c} and Supplementary Movie 2). 
%
%
It is noteworthy that the magnitude of the induced magnetization is
only an order of magnitude smaller than in the induced (scalar) charge (Fig.~{\ref{fig:K_mod03}}{\bf b}). 
%
%
Since ${\bf q}$=${\bf K'}$ is the time-reversed momentum of ${\bf q}$=${\bf K}$, 
as mentioned earlier, the real-space magnetization 
should be  opposite in sign.
For the first example given above but in the case in which ${\bf q}$=${\bf K'}$,  Se atoms rotate 
anticlockwise in the plane and the W vibrates perpendicular to the plane (Supplementary Movie 3). 
Taking a snapshot in time for which the atomic positions coincide with those in Fig.~{\ref{fig:K_mod02}
 demonstrates that the magnetization shows exactly opposite chirality, and therefore
 the relation $\delta{\bf M}_{{\bf K},\nu}({\bf r}, t)=-\delta{\bf M}_{{\bf -K},\nu}({\bf r}, t)$
is fulfilled  (Supplementary Figure~1).
Note that the propagation of the atomic displacements of W in the perpendicular direction  to the plane 
along ${\bf q}$=${\bf K}$  (Fig.~{\ref{fig:K_mod02}{\bf b})
 is exactly the same as the one along ${\bf q}$=${\bf K'}$ 
when it is looked from right to left (Supplementary Figure~1{\bf b}).
A similar situation occurs for the second example, where the chirality of the atomic displacements reverses
  when changing ${\bf q}$=${\bf K}$ to its time-reversed value 
\cite{chiralphonons2018}
 which again gives a magnetization opposite in sign
(Supplementary Movie 4)
 and which can be compared to 
that of Fig.~{\ref{fig:K_mod03} when the atomic positions are frozen to be the same 
(Supplementary Figure 2). 
%
 The non-collinear character in space and time of the magnetization is also observed when 
the atomic displacements are all linear. For instance, for an acoustic phonon with vector
 ${\bf q}$=${\bf M}$  at which Se atoms move along 
${\bf q}$ and W atoms vibrate in the perpendicular direction to the plane (Supplementary Movie 5) a chiral magnetization pattern similar to that
 of Fig.~{\ref{fig:K_mod02} is found  (Supplementary Figure~3). 
Instead, when all atoms oscillate linearly in the plane at right angles to the propagation vector ${\bf q}$=${\bf M}$ (Supplementary Movie 6), the magnetization appears in the perpendicular direction to the plane
as in  Fig.~{\ref{fig:K_mod03} (Supplementary Figure~4). 
Note that for this acoustic mode when all atoms go through their equilibrium positions both, the induced charge and 
spin-density fields, disappear.
All the magnetization patterns show a periodicity in real space
according to the  wave number ${\bf q}$ of the propagation of the excited phonon, as they should.
At this point it is worth mentioning that the magnetic polarization of the
electron gas as described in this manuscript does not have a relation with the
angular momentum of the atoms as described by Zhang et al. in \cite{Niu2014}. In our theory 
linearly polarized phonons with null angular momentum give a finite and meaningful contribution to the magnetization. It 
is therefore clear that the physics described in \cite{Niu2014} is different and not connected to the 
spin response of the electron gas as described in the present work.

{The size of the fluctuations of the real space unit-cell average of these 
oscillations gives an order of the magnitude  of this effect, even though it does not capture all the
 detailed structure in real space. }
 It is nevertheless
 physically meaningful, allowing to analyze the momentum and mode dependence at the same time,
and making a connection with the possibility of experimental detection as it will be shown shortly.
More specifically, {for a given phonon $\mathbf{q}$} 
the root-mean-square (RMS) of the time dependence of the {periodic part} of this quantity reflects the overall
amplitude of the spin-density associated to a single phonon mode.
For each cartesian component $\alpha$ we have (see Supplementary Note 2): 
\begin{eqnarray} \label{eq:Mcompfluct}
	\overline{\delta M^{\alpha}_{{\bf q},\nu}} = \sqrt{\left\langle  \left ( {\Re} \int_{\Omega} \frac{d^3{\bf r}}{\Omega}  \left [
		\delta  \tilde{M}^{\alpha}_{{\bf q},\nu} ({\bf r}) e^{-i \omega_{{\bf q},\nu} t} \right ] \right )^2 \right\rangle_{T}} .
\end{eqnarray}
The above classical RMS amplitudes of the magnetization are directly connected to the 
charge-charge, spin-charge and spin-spin components of the
 dynamic structure factor,  $S_{\alpha,\beta}(\omega, {\bf q+G})$ 
{(see Supplementary Note 2)},
 which is accessed by 
inelastic neutron scattering, inelastic X-Ray spectroscopy and spin-polarized electron energy loss spectroscopy \cite{sturm93}. 
As van Hove first pointed out \cite{vanHove}, 
the dynamic structure factor is the space and time Fourier transform of the density-density
correlation function
 $\langle \delta \hat{{\bf n}}^{\alpha}({\bf r},t)\delta \hat{{\bf n}}^{\beta}({\bf r'},0) \rangle_T$.
 If we consider a 4-dimensional spin-charge quantized field %
\begin{eqnarray} \label{eq:nfield}
\delta \hat{{\bf n}}({\bf r},t) = \sum_{{\bf q}\nu}
\left ( a^{\dagger}_{{\bf q},\nu} e^{i \omega_{{\bf q},\nu} t} \delta{\bf \tilde{n}}^{*}_{{\bf q},\nu}({\bf r})  + 
a_{{\bf q},\nu} e^{-i \omega_{{\bf q},\nu} t} \delta{\bf \tilde{n}}_{{\bf q},\nu}({\bf r}) \right  ),
\end{eqnarray}
where  $a_{{\bf q},\nu}$ and $a^{\dagger}_{{\bf q},\nu}$ are creation and annihilation operators for 
a phonon mode (${\bf q},\nu$) with energy $\omega_{{\bf q},\nu}$ {(Fig.~\ref{fig:ph_mag}{\bf a})}, then the  dynamic structure factor
can be written as
\begin{eqnarray} \label{eq:Sqw}
S_{\alpha,\beta}(\omega, {\bf q+G})&=&
[1+f_B(\omega_{{\bf q},\nu })] \delta\tilde{\tilde{n}}_{{\bf q},\nu}^{\alpha}({\bf G}) \delta\tilde{\tilde{n}}_{{\bf q},\nu}^{\beta *}({\bf G})  
\delta (\omega - \omega_{{\bf q},\nu }) 
\\
\nonumber 
&+& f_B(\omega_{{\bf q},\nu }) \delta\tilde{\tilde{n}}_{{\bf q},\nu}^{\alpha *}({\bf G}) \delta\tilde{\tilde{n}}_{{\bf q},\nu}^{\beta}({\bf G}) 
\delta (\omega + \omega_{{\bf q},\nu }),
\end{eqnarray} 
where $\delta\tilde{\tilde{n}}_{{\bf q},\nu}^{\alpha}({\bf G})$ indicates the Fourier transform or crystal field components of the real-space 
complex amplitudes $\delta\tilde{n}_{{\bf q},\nu}^{\alpha}({\bf r})$,
 $f_B(\omega_{{\bf q},\nu })$ denotes phonon occupation numbers and where we ignore the Debye-Waller factor \cite{Jancovici67}.
Taking the ${\bf G=0}$ components (unit-cell average), it is easily seen 
that the classical RMS terms of equation~(\ref{eq:Mcompfluct})
are proportional to the diagonal ($\alpha=\beta$)
spectral contributions to equation (\ref{eq:Sqw}) for individual phonons: 
$\left(\overline{\delta M^{\alpha}_{{\bf q},\nu}}\right)^2 \sim 
\delta\tilde{\tilde{n}}_{{\bf q},\nu}^{\alpha *}({\bf 0}) \delta\tilde{\tilde{n}}_{{\bf q},\nu}^{\alpha}({\bf 0})$. 
This helps to physically interpret the RMS of the 
induced magnetization defined as above because the 
terms parallel
 ($\sqrt{(\overline{\delta M^{x}_{{\bf q},\nu}})^2+(\overline{\delta M^{y}_{{\bf q},\nu}})^2}$)
and  perpendicular ($\overline{\delta M^{z}_{{\bf q},\nu}}$)   to
the WSe$_2$ layer shown in Fig.~\ref{fig:ph_mag}{\bf b} and Fig.~\ref{fig:ph_mag}{\bf c}, respectively,
are basically the spin contributions to the structure factor 
connected to a given phonon mode $({\bf q},\nu)$.
Said in other words, Fig.~\ref{fig:ph_mag}{\bf b} and Fig.~\ref{fig:ph_mag}{\bf c}  may be interpreted as the momentum/energy and phonon mode
resolved contributions to the spin sector of the dynamic structure factor depicted along the high symmetry lines
{of the surface Brillouin zone}. 
The magnetic character associated inherently to phonons as proposed in this work is therefore accessible by means of any 
experimental setup probing the spin components of the dynamical structure factor in the energy ranges corresponding to phonons.
%

\section*{Discussion}
We conclude that in any material with a non-trivial spin-pattern within the Brillouin zone, even if non-magnetic,
phonon modes are connected inherently to a magnetic property 
analogous to the electron spin-polarization
and it  can be stated quite generally that phonon modes 
are accompanied by an induced spin-density (magnetization) which is rich in real space details.
It is also shown  that  this magnetic modulation is only one
order of magnitude weaker than the purely electrostatic (spin-diagonal) terms.
All the above physics is illustrated convincingly for WSe$_2$ in a mode by mode analysis where
the real space and time dependence of the induced magnetization is revealed for the 
most relevant modes. 
The implications are extensive and profound because phonons, which are now intrinsically
attached  to an effective magnetic moment, 
should be understood as an additional spin-flip mechanism even for materials without a net magnetic moment.   
This means  that the whole electron-phonon physics is modulated 
in every system with spin-orbit coupling and, for instance,
even electron backscattering events may be aided by the phonon magnetic moment. 
Experimental detection of magnetic oscillations for coherent phonons 
should be done by ultrafast probes and our calculated
details of these fields may indicate a detection strategy.
However, we also show that 
probing the spin components of the dynamical structure factor may be an alternative route to 
measure what we could name as the spin polarization of phonons.

\begin{addendum}
 \item[Data availability]
 The data that support the findings of this study are available from the corresponding
author upon reasonable request.
\end{addendum}


\begin{addendum}
 \item [Acknowledgements]
The authors acknowledge the
Department of Education, Universities and Research of the
Basque Government and UPV/EHU (Grant No. IT756-13),
the Spanish Ministry of Economy and Competitiveness MINECO 
(Grant No.  FIS2016-75862-P) and the University of the Basque Country UPV/EHU 
(Grant No. GIU18/138) for financial support. 
Computer facilities
were provided by the Donostia International Physics Center (DIPC).

\item [Author contributions]
A.E. conceived the ideas. A.I. contributed to the early stages of the calculations.
I.G.G. and A.E. carried out the calculations, discussed the results and contributed to the writing of the manuscript.

\item [Competing interests]
The authors declare no competing interests.

\end{addendum}

\begin{figure*}
\begin{center}
\includegraphics[width=1.0\textwidth]{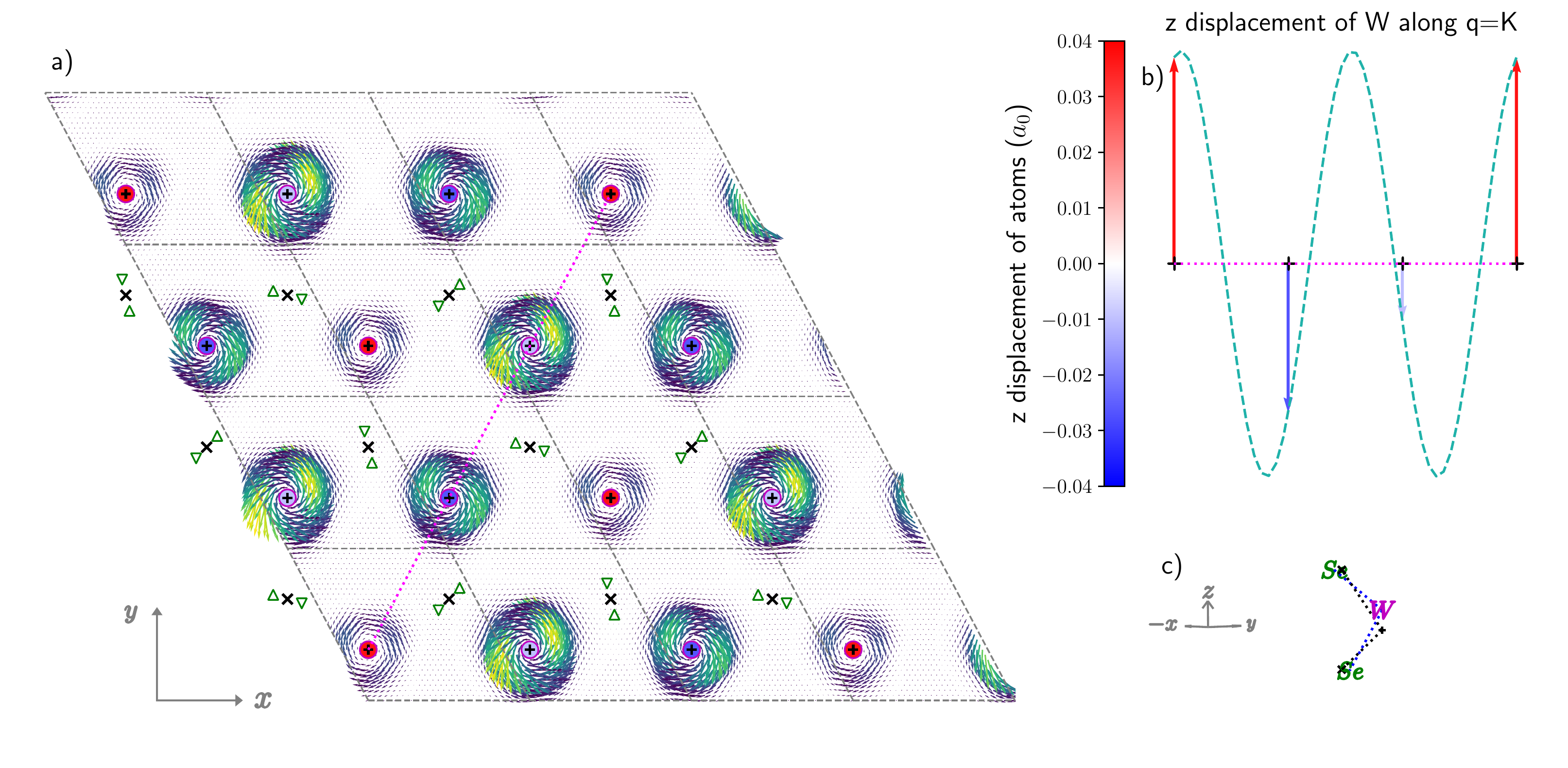}
\end{center}
\caption{{\bf 
Induced magnetization for the second lowest energy acoustic phonon
mode at ${\bf q}={\bf K}$ involving pure out of plane displacements of W atoms }
{\bf a)} Real-space representation of the magnetization in the plane of the W atoms
for $4\times4$ unit cells 
along the hexagonal axes of WSe$_2$ for 
the second lowest energy mode at  ${\bf q}={\bf K}$ (in orange
 in Figs. ~\ref{fig:ph_mag}{\bf a} and ~\ref{fig:ph_mag}{\bf b}).
In this mode the W atoms (filled circles) 
displace along the 
perpendicular direction  (see colour-bar)
 and the Se atoms above (filled triangle up) and below (filled triangle down) 
the W plane rotate clockwise with opposite phase around their equilibrium positions (crosses) in their respective planes.
The coloured vector-field is proportional to the in-plane magnetization at each point in real space,
with yellow/light (blue/dark) arrows 
representing the largest (smallest)  values. 
These arrows as well as the displacements of the Se atoms have been scaled to make them visible. 
{\bf b)} The coloured arrows give the z displacement of the W atoms along the ${\bf q}={\bf K}$ direction
 (dotted  magenta line in panel {\bf a)}) according to the  colour-bar. 
The dashed line describes the propagation of the vibration  along several unit-cells in real space.
Note that ${\bf K}=[1/3, 1/3, 0]$ in crystal axes, and hence the periodicity of the wave.
{\bf c)} Side view of the WSe$_2$ formula-unit in the lower left corner unit-cell. 
The names of the atoms display their
 displacements from the equilibrium positions, denoted as in panel {\bf a)}.
This figure is a snapshot of the time evolution of the induced magnetization for this mode
 (Supplementary Movie 1).
}
\label{fig:K_mod02}
\end{figure*}

\begin{figure*}
\begin{center}
\includegraphics[width=1.0\textwidth]{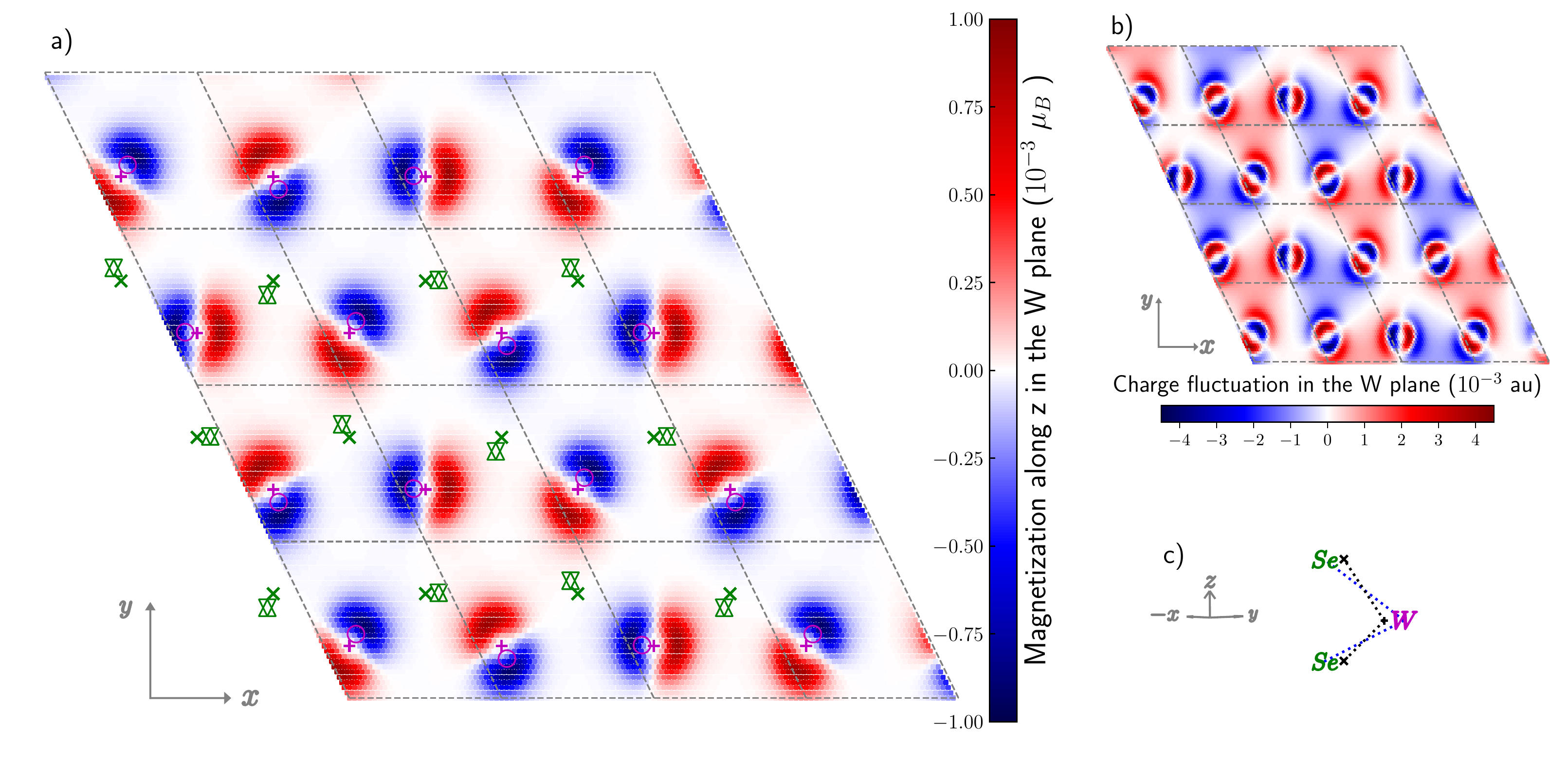}
\end{center}
\caption{
{\bf 
Induced magnetization for the highest energy acoustic phonon
mode at ${\bf q}={\bf K}$ involving in-plane displacements of the W atoms}
 {\bf a)} 
Real-space representation of the perpendicular component of the magnetization in $4\times4$ unit cells along the hexagonal axes of WSe$_2$
for the highest energy  acoustic mode for ${\bf q}={\bf K}$ 
(represented in green in Figs.~\ref{fig:ph_mag}{\bf a} and ~\ref{fig:ph_mag}{\bf c}).
This mode is composed by clockwise rotations of the W atoms (circles)
 around their equilibrium positions 
 and in-phase anticlockwise rotations of the Se atoms located above (triangle up)
and below (triangle down) the W plane. 
The colour code represents the magnetization in the perpendicular direction.
The displacements of the atoms have been scaled to make them visible.
{\bf b)} Induced electronic charge  for the same atomic
 configuration as in panel  {\bf a)}. 
{Note that the induced magnetization is only an order of magnitude smaller than the induced
(scalar) charge.}
{\bf c)} Side view of the WSe$_2$ formula-unit in the lower left corner unit-cell. 
The names of the atoms display their
 displacements from the equilibrium positions, denoted as in panel {\bf a)}.
This figure is a snapshot of the time evolution of the induced magnetization
 for this mode (Supplementary Movie 2).
}
\label{fig:K_mod03}
\end{figure*}

\begin{figure*}
\begin{center}
\includegraphics[width=0.75\textwidth]{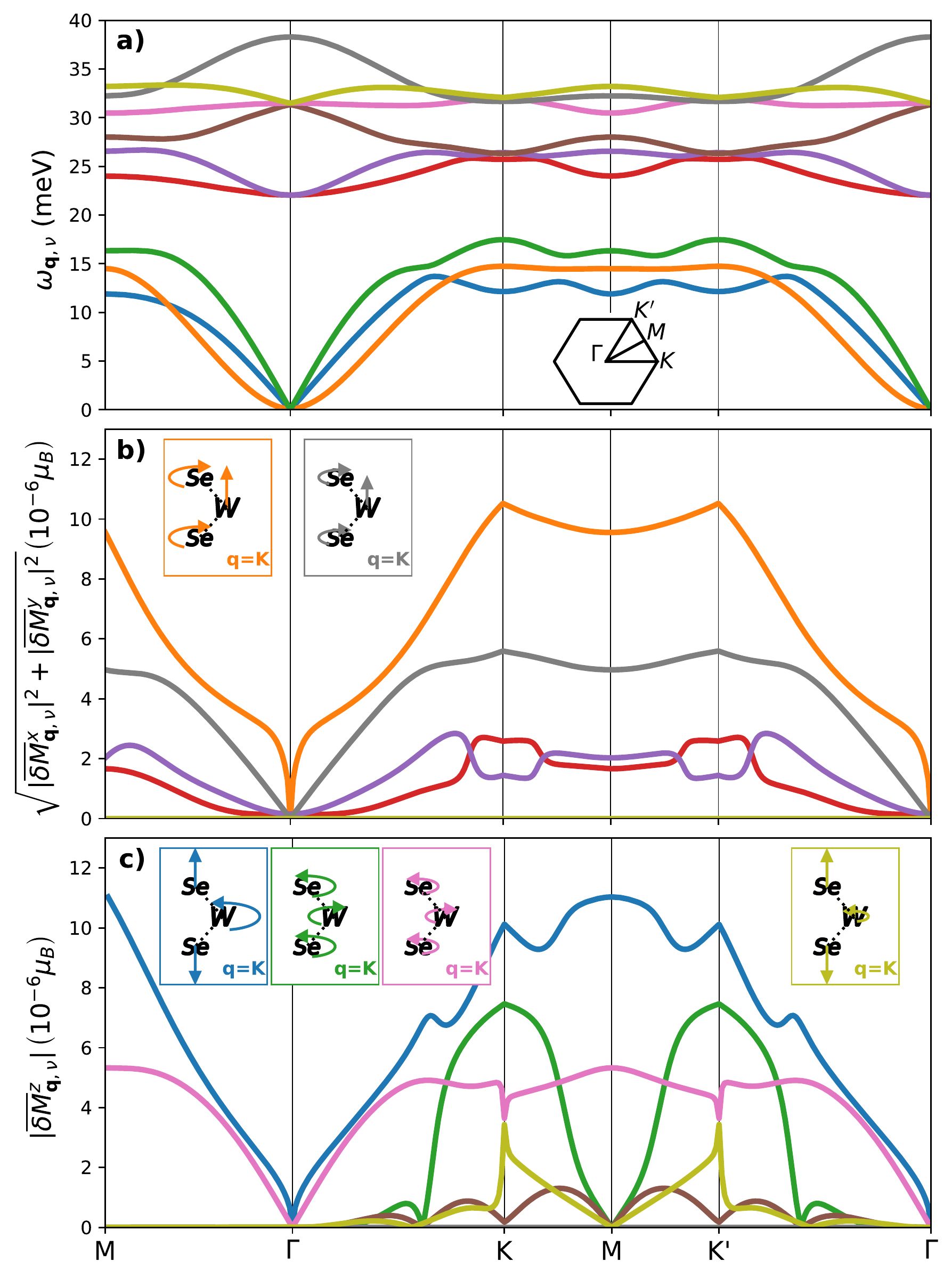}
\end{center}
\caption{
{\bf 
Mode and momentum resolved magnetization induced by lattice vibrations in monolayer WSe$_2$}
{\bf a)} Phonon energy spectrum along high symmetry lines in the surface Brillouin zone (inset).
{\bf b)} and {\bf c)} Plane {($x$, $y$)} and out-of-plane {($z$)} components of the magnitude of the 
 unit-cell average magnetization 
for each phonon mode in {\bf a)} and with the same colour convention. 
Insets show the corresponding polarization vectors for ${\bf q}={\bf K}$, 
the length of the arrows being proportional 
to the magnitude of the atomic displacements. {Vertical arrows represent linear displacements in the perpendicular direction to the plane, and semicircular arrows show circular displacements of the atoms 
around their equilibrium positions.}
The direction of the magnetization is 
determined by the motion of the W atom.
When W vibrates in the perpendicular direction of the plane, it induces
a net magnetization in the plane (panel {\bf b)}).
However, when W atoms move  on the plane (with circular polarization for ${\bf q=K}$), 
the induced magnetization appears in the perpendicular direction (panel {\bf c)}). 
For the three middle modes in the phonon spectrum W atoms move significantly less, 
yielding a smaller magnetization.
Panels {\bf b)} and {\bf c)} can be interpreted as a frequency and momentum resolved
 dynamic structure factor of phonons. 
}
\label{fig:ph_mag}
\end{figure*}


\end{document}